# Constructive Dimension and Turing Degrees

Laurent Bienvenu*         David Doty†         Frank Stephan‡
Université de Provence    Iowa State University    National University of Singapore

## Abstract

This paper examines the constructive Hausdorff and packing dimensions of Turing degrees. The main result is that every infinite sequence $S$ with constructive Hausdorff dimension $\dim_H(S)$ and constructive packing dimension $\dim_P(S)$ is Turing equivalent to a sequence $R$ with $\dim_H(R) \geq (\dim_H(S)/\dim_P(S)) - \epsilon$, for arbitrary $\epsilon > 0$. Furthermore, if $\dim_P(S) > 0$, then $\dim_P(R) \geq 1 - \epsilon$. The reduction thus serves as a *randomness extractor* that increases the algorithmic randomness of $S$, as measured by constructive dimension.

A number of applications of this result shed new light on the constructive dimensions of Turing degrees. A lower bound of $\dim_H(S)/\dim_P(S)$ is shown to hold for the Turing degree of any sequence $S$. A new proof is given of a previously-known zero-one law for the constructive packing dimension of Turing degrees. It is also shown that, for any *regular* sequence $S$ (that is, $\dim_H(S) = \dim_P(S)$) such that $\dim_H(S) > 0$, the Turing degree of $S$ has constructive Hausdorff and packing dimension equal to 1.

Finally, it is shown that no single Turing reduction can be a *universal* constructive Hausdorff dimension extractor, and that *bounded* Turing reductions cannot extract constructive Hausdorff dimension. We also exhibit sequences on which weak truth-table and bounded Turing reductions differ in their ability to extract dimension.

## 1 Introduction

Hausdorff [8] initiated the study of dimension as a general framework to define the size of subsets of metric spaces. Recently this framework had been effectivized; Lutz [13] gives an overview of this historical development. Furthermore, Lutz [12, Section 6] reviews early results that anticipated the effectivization of Hausdorff dimension. *Constructive Hausdorff dimension* was defined by Lutz [12] to study effective dimension at the level of computability theory. Intuitively, given an infinite binary sequence $S$ – interpreted as a language or decision problem – the constructive Hausdorff dimension $\dim_H(S)$ of $S$ is a real number in the interval [0,1] indicating the density of algorithmic randomness of the sequence. The constructive Hausdorff dimension of a class $\mathcal{C}$ of sequences is the supremum of the dimensions of individual sequences in $\mathcal{C}$. For many classes $\mathcal{C}$ of interest in computability theory, the problem of determining the constructive Hausdorff dimension of $\mathcal{C}$ remains open.

*Laboratoire d'Informatique Fondamentale de Marseille, Université de Provence, 39 rue Joliot-Curie, 13453 Marseille Cedex 13, France. laurent.bienvenu@lif.univ-mrs.fr.
†Department of Computer Science, Iowa State University, Ames, IA 50011, USA. ddoty@iastate.edu.
‡School of Computing and Department of Mathematics, National University of Singapore, 2 Science Drive 2, Singapore 117543, Republic of Singapore. fstephan@comp.nus.edu.sg. Supported in part by NUS research grants no. R252-000-212-112 and R252-000-308-112.

Independently of each other, Reimann [20] and Terwijn investigated in particular whether there are degrees of fractional constructive Hausdorff dimension. Stated in terms of individual sequences, Reimann and Terwijn asked which reducibilities (such as Turing, many-one, weak truth-table etc.) are capable of increasing the constructive Hausdorff dimension of a sequence. We call such a reduction a *dimension extractor*, since its purpose bears a resemblance to that of the *randomness extractors* of computational complexity [24], which are algorithms that turn a source of weak randomness (a probabilistic source with low entropy) into a source of strong randomness (a source with high entropy). Viewing a sequence with positive, but still fractional, constructive Hausdorff dimension as a weak source of randomness, Reimann essentially asked whether such randomness can be extracted via a reduction to create a sequence with dimension closer to 1. If such extraction is *not* possible for some sequence $S$, this indicates that the degree of $S$ under the reduction has fractional dimension.

A number of negative results for dimension extractors are known. Reimann [20, Theorem 3.10] and Terwijn proved that there are many-one and bounded truth-table degrees with constructive Hausdorff dimension strictly between 0 and 1. Later Reimann and Slaman [21] extended this result to truth-table degrees. Stephan [26] showed that there is a relativized world in which there exists a wtt degree of constructive Hausdorff dimension between $\frac{1}{4}$ and $\frac{1}{2}$. Furthermore, Nies and Reimann [16] obtained a non-relativized variant of this result and constructed, for each rational $\alpha$ between 0 and 1, a wtt degree of constructive Hausdorff dimension $\alpha$.

Doty [5] attempted positive results by considering the interaction between constructive Hausdorff dimension and *constructive packing dimension* [1], a dual quantity that is a constructive effectivization of classical packing dimension [27, 28], another widely-studied fractal dimension. The constructive packing dimension $\dim_P(S)$ of a sequence $S$ always obeys

$$0 \leq \dim_H(S) \leq \dim_P(S) \leq 1,$$

with each inequality tight in the strong sense that there are sequences $S$ in which $\dim_H(S)$ and $\dim_P(S)$ may take on any values obeying the stated constraint. Doty showed that every sequence $S$ with $\dim_H(S) > 0$ is Turing equivalent to a sequence $R$ with $\dim_P(R) \geq 1 - \epsilon$, for arbitrary $\epsilon > 0$. This implies that the constructive packing dimension of the Turing degree of any sequence $S$ with $\dim_H(S) > 0$ is equal to 1. Unfortunately, since $\dim_H(R) \leq \dim_P(R)$, this Turing reduction constitutes a weaker example of a dimension extractor than that sought by Reimann and it tells us nothing of the constructive dimensions of arbitrary Turing degrees.

Doty [5] obtained stronger results for other effective dimensions such as computable dimension and various time and space bounded dimensions [11]. Zimand [30] has shown that, given two *independent* sequences with positive constructive Hausdorff dimension, a truth-table reduction with access to both sequences suffices to compute a sequence with constructive Hausdorff dimension equal to 1.

We obtain in the current paper stronger positive results for constructive dimension extractors, dual to those obtained for other effective dimensions in [5]. Our main result, in section 2, is that, given any infinite sequence $S$ and $\epsilon > 0$, there exists $R \equiv_T S$ such that $\dim_H(R) \geq \frac{\dim_H(S)}{\dim_P(S)} - \epsilon$ and, if $\dim_P(S) > 0$, then $\dim_P(R) \geq 1 - \epsilon$. This has immediate consequences for the dimensions of Turing degrees:

- Given any sequence $S$, $\dim_H(\deg_T(S)) \geq \frac{\dim_H(S)}{\dim_P(S)}$.

- If $\dim_P(S) > 0$, then $\dim_P(\deg_T(S)) = 1$, implying that *every* Turing degree has constructive packing dimension 0 or 1.

- Given any *regular* sequence $S$ such that $\dim_H(S) > 0$, $\dim_H(\deg_T(S)) = 1$, where a sequence $S$ is called regular if it satisfies $\dim_H(S) = \dim_P(S)$.

In section 3, we use Theorem 2.1 to show that, for every $\alpha > 0$, there is no *universal* Turing reduction that is guaranteed to extract dimension from all sequences of dimension at least $\alpha$. We also obtain the result that *bounded* Turing reductions (which are allowed to make at most a constant number of queries to the input sequence) cannot extract dimension, and we show examples of sequences in which bounded reductions and Turing reductions differ in their ability to extract dimension. In particular, we show that there are sequences from which some bounded reduction can extract packing dimension 1, while no Turing reduction can compute a sequence with packing dimension greater than 0, and there are sequences from which some wtt reduction can extract *Hausdorff* dimension 1, while no bounded reduction can compute a sequence with packing dimension greater than 0.

Very recently, Joe Miller [15] answered Reimann's original question by extending the result of Reimann and Nies from wtt reductions to Turing reductions; that is, there are sequences of positive constructive Hausdorff dimension that compute (via Turing reductions) no sequence of higher constructive Hausdorff dimension. This result provides as a corollary an independent proof of Theorem 4.1 of the current paper. Combined with Zimand's previously mentioned two-source extractor [30], this mirrors a phenomenon in randomness extractors used in computational complexity, which is that no function can extract randomness from a single classical probabilistic source; two or more *independent* sources of randomness are required. See [24] for an explanation of this phenomenon and a survey of randomness extractors in computational complexity.

This paper corrects an incorrect proof of Theorem 2.4 in [2], a preliminary version of the current paper. In that paper, the incorrect proof of Theorem 2.4 implied that the reduction used was a weak truth-table reduction. The corrected version of the proof, of Theorem 2.4 in the current paper, however, uses a Turing reduction that is not weak truth-table. Hence the conclusions about weak truth-table degrees in Section 2 of [2] are known only to hold for Turing degrees. Furthermore, the printed version of this paper appearing in *Theory of Computing Systems*, 45(4):740-755, 2009, had another error in the proof of Theorem 2.4, due to insufficient care with the choice of $\delta$. This version modifies that proof to fix the error.

Before going into the details of the results, we introduce the concepts and notations formally.

**Notation.** We refer the reader to the textbooks of Li and Vitányi [10] for an introduction to Kolmogorov complexity and algorithmic information theory and of Odifreddi [18] and Soare [25] for an introduction to computability theory. Although we follow mainly the notation in these books, we nevertheless want to remind the reader of the following definitions, either for the reader's convenience or because we had to choose between several common ways of denoting the corresponding mathematical objects.

All logarithms are base 2. $\mathbb{N}$ denotes the set $\{0, 1, 2, 3, \ldots\}$ of the natural numbers including 0. $\{0,1\}^*$ denotes the set of all finite, binary *strings*. For all $x \in \{0,1\}^*$, $|x|$ denotes the *length* of $x$. $\lambda$ denotes the empty string. $\mathbf{C} = \{0,1\}^\infty$ denotes the *Cantor space*, the set of all infinite, binary *sequences*. For $x \in \{0,1\}^*$ and $y \in \{0,1\}^* \cup \mathbf{C}$, $xy$ denotes the concatenation of $x$ and $y$, $x \sqsubseteq y$ denotes that $x$ is a *prefix* of $y$ (that is, there exists $u \in \{0,1\}^* \cup \mathbf{C}$ such that $xu = y$) and $x \sqsubset y$ denotes that $x \sqsubseteq y$ and $x \neq y$. For $S \in \{0,1\}^* \cup \mathbf{C}$ and $i, j \in \mathbb{N}$, $S[i]$ denotes the $i^{\text{th}}$ bit of $S$, with $S[0]$ being the leftmost bit, $S[i\mathinner{.\,.}j]$ denotes the substring consisting of the $i^{\text{th}}$ through $j^{\text{th}}$ bits of $S$ (inclusive), with $S[i\mathinner{.\,.}j] = \lambda$ if $i > j$.

**Reductions and Compression.** Let $M$ be a Turing machine and $S \in \mathbf{C}$. We say $M$ *computes* $S$ if $M$ on input $n \in \mathbb{N}$ (written $M(n)$), outputs the string $S[0\mathinner{.\,.}n-1]$. We define an *oracle Turing machine* to be a Turing machine $M$ that can make constant-time queries to an oracle sequence and we let OTM denote the set of all oracle Turing machines [29]. For $R \in \mathbf{C}$, we say $M$ operates *with oracle* $R$ if, whenever $M$ makes a query to index $n \in \mathbb{N}$, the bit $R[n]$ is returned. We write $M^R$ to denote the oracle Turing machine $M$ with oracle $R$. We identify an infinite binary sequence $S \in \mathbf{C}$ with the language of which $S$ is the characteristic sequence; that is, the set that contains $n$ exactly when $S[n] = 1$. In such a case, given $x \in \{0,1\}^*$, $S[x]$ may also be interpreted as the binary condition "$x$ is in the language $S$". Similarly, we identify each infinite binary sequence as representing the characteristic sequence of a subset of $\mathbb{N}$.

Let $S, R \in \mathbf{C}$ and $M \in \text{OTM}$. We say $S$ *is Turing reducible to* $R$ *via* $M$ and we write $S \leq_{\text{T}} R$ *via* $M$, if $M^R$ computes $S$ (that is, if $M^R(n) = S[0\mathinner{.\,.}n-1]$ for all $n \in \mathbb{N}$). In this case, write $R = M(S)$. We say $S$ *is Turing reducible to* $R$ and we write $S \leq_{\text{T}} R$, if there exists $M \in \text{OTM}$ such that $S \leq_{\text{T}} R$ via $M$. We say $S$ is *Turing equivalent* to $R$, and we write $S \equiv_{\text{T}} R$, if $S \leq_{\text{T}} R$ and $R \leq_{\text{T}} S$. The *Turing lower span* of $S$ is $\text{span}_{\text{T}}(S) = \{\, R \in \mathbf{C} \mid R \leq_{\text{T}} S \,\}$ and the *Turing degree* of $S$ is $\deg_{\text{T}}(S) = \{\, R \in \mathbf{C} \mid R \equiv_{\text{T}} S \,\}$.

Let $S, R \in \mathbf{C}$ and $M \in \text{OTM}$ such that $S \leq_{\text{T}} R$ via the oracle Turing machine $M$. Let $\#(M^R, S[0\mathinner{.\,.}n-1])$ denote the *query usage of* $M^R$ *on* $S[0\mathinner{.\,.}n-1]$, the index of the rightmost bit of $R$ queried by $M$ when computing $S[0\mathinner{\ldotp\ldotp} n-1]$.

We say *$S$ is weak truth-table (wtt) reducible to $R$ via $M$* and we write $S \leq_{\mathrm{wtt}} R$ via $M$, if $S \leq_{\mathrm{T}} R$ via $M$ and there is a computable function $q : \mathbb{N} \to \mathbb{N}$ such that, for all $n \in \mathbb{N}$, $\#(M^R, S[0\mathinner{\ldotp\ldotp} n-1]) \leq q(n)$ (see [7]).

We say *$S$ is bounded Turing (bT) reducible to $R$ via $M$* (or simply *bounded reducible*) and we write $S \leq_{\mathrm{bT}} R$ via $M$, if $S \leq_{\mathrm{T}} R$ via $M$ and there exists a constant $c \in \mathbb{N}$ such that, for every $n \in \mathbb{N}$, the number of bits of $R$ queried when computing $S[n]$ is at most $c$ (see [9, 18]).

We say *$S$ is truth-table (tt) reducible to $R$ via $M$* and we write $S \leq_{\mathrm{tt}} R$ via $M$, if $S \leq_{\mathrm{T}} R$ via $M$ and, for all $R' \in \mathbf{C}$, $M(R')$ is defined; that is, $M$ is total with respect to the oracle (see [19]).

Bounded reductions and wtt reductions are incomparable. However, every tt reduction is a wtt reduction.

Given $\Delta \in \{\mathrm{tt}, \mathrm{wtt}, \mathrm{bT}\}$, define $S \leq_\Delta R$, $S \equiv_\Delta R$, $\mathrm{span}_\Delta(S)$ and $\deg_\Delta(S)$ analogously to their counterparts for Turing reductions.

Define

$$\begin{aligned}
\rho_M^-(S, R) &= \liminf_{n \to \infty} \frac{\#(M^R, S[0\mathinner{\ldotp\ldotp} n-1])}{n}, \\
\rho_M^+(S, R) &= \limsup_{n \to \infty} \frac{\#(M^R, S[0\mathinner{\ldotp\ldotp} n-1])}{n}.
\end{aligned}$$

Viewing $R$ as a compressed version of $S$, $\rho_M^-(S, R)$ and $\rho_M^+(S, R)$ are respectively the best- and worst-case compression ratios as $M$ decompresses $R$ into $S$. Note that $0 \leq \rho_M^-(S, R) \leq \rho_M^+(S, R) \leq \infty$.

The following lemma is useful when one wants to compose two reductions:

**Lemma 1.1** (Doty [4]). *Let $S, S', S'' \in \mathbf{C}$ and $M_1, M_2 \in \mathrm{OTM}$ such that $S' \leq_{\mathrm{T}} S$ via $M_1$ and $S'' \leq_{\mathrm{T}} S'$ via $M_2$. There exists $M \in \mathrm{OTM}$ such that $S'' \leq_{\mathrm{T}} S$ via $M$ and:*

$$\begin{aligned}
\rho_M^+(S'', S) &\leq \rho_{M_2}^+(S'', S')\rho_{M_1}^+(S', S). \\
\rho_M^-(S'', S) &\leq \rho_{M_2}^-(S'', S')\rho_{M_1}^+(S', S). \\
\rho_M^-(S'', S) &\leq \rho_{M_2}^+(S'', S')\rho_{M_1}^-(S', S).
\end{aligned}$$

(The last bound is not explicitly stated in [4], but it holds for the same reason as the second one).

For $S \in \mathbf{C}$, the *lower and upper Turing compression ratios of $S$* are respectively defined as

$$\begin{aligned}
\rho^-(S) &= \min_{\substack{R \in \mathbf{C} \\ M \in \mathrm{OTM}}} \left\{ \rho_M^-(S, R) \;\middle|\; S \leq_{\mathrm{T}} R \text{ via } M \right\}, \\
\rho^+(S) &= \min_{\substack{R \in \mathbf{C} \\ M \in \mathrm{OTM}}} \left\{ \rho_M^+(S, R) \;\middle|\; S \leq_{\mathrm{T}} R \text{ via } M \right\}.
\end{aligned}$$

Doty [4] showed that the above minima exist. Note that $0 \leq \rho^-(S) \leq \rho^+(S) \leq 1$.

**Constructive Dimension.** Lutz [12] gives an introduction to the theory of constructive dimension. We use Mayordomo's characterization [14] of the constructive dimensions of sequences. For all $S \in \mathbf{C}$, the *constructive Hausdorff dimension* and the *constructive packing dimension* of $S$ are respectively defined as

$$\dim_{\mathrm{H}}(S) = \liminf_{n \to \infty} \frac{\mathrm{C}(S[0..n-1])}{n} \text{ and } \dim_{\mathrm{P}}(S) = \limsup_{n \to \infty} \frac{\mathrm{C}(S[0..n-1])}{n},$$

where $\mathrm{C}(w)$ denotes the *Kolmogorov complexity* of $w \in \{0,1\}^*$ (see [10]). If $\dim_{\mathrm{H}}(S) = \dim_{\mathrm{P}}(S)$, we say $S$ is a *regular* sequence. Doty [4] showed that, for all $S \in \mathbf{C}$, $\rho^-(S) = \dim_{\mathrm{H}}(S)$ and $\rho^+(S) = \dim_{\mathrm{P}}(S)$.

For all $X \subseteq \mathbf{C}$, the *constructive Hausdorff dimension* and the *constructive packing dimension* of $X$ are respectively defined as

$$\dim_{\mathrm{H}}(X) = \sup_{S \in X} \dim_{\mathrm{H}}(S) \text{ and } \dim_{\mathrm{P}}(X) = \sup_{S \in X} \dim_{\mathrm{P}}(S).$$

## 2 Constructive Dimension Extractors

Nies and Reimann [16] showed that wtt reductions cannot always extract constructive dimension.

**Theorem 2.1** (Nies and Reimann [16]). *For every rational number $\alpha$ with $0 < \alpha < 1$, there exists a sequence $S \in \mathbf{C}$ such that, for all wtt reductions $M$, $\dim_{\mathrm{H}}(M(S)) \leq \dim_{\mathrm{H}}(S) = \alpha$.*

Ryabko [22, 23] discovered the next theorem.

**Theorem 2.2** (Ryabko [22, 23]). *For all $S \in \mathbf{C}$ and $\delta > 0$, there exists $R \in \mathbf{C}$ and $N_d \in \mathrm{OTM}$ such that*

1. *$S \leq_{\mathrm{T}} R$ via $N_d$ and $R \leq_{\mathrm{T}} S$.*

2. *$\rho^-_{N_d}(S, R) \leq \dim_{\mathrm{H}}(S) + \delta$.*

The following theorem was shown in [4].

**Theorem 2.3** (Doty [4]). *There is an oracle Turing machine $M_d$ such that, for all $S \in \mathbf{C}$, there exists $R \in \mathbf{C}$ such that*

1. *$S \leq_{\mathrm{wtt}} R$ via $M_d$.*

2. *$\rho^-_{M_d}(S, R) = \dim_{\mathrm{H}}(S)$.*

3. *$\rho^+_{M_d}(S, R) = \dim_{\mathrm{P}}(S)$.*

The following theorem, which is similar to Ryabko's Theorem 2.2, shows that the decoding machine $M_d$ of Theorem 2.3 can also be reversed if the compression requirements are weakened, and if the compression direction is allowed to be a Turing, rather than a weak truth-table, reduction.

**Theorem 2.4.** *Let $M_d$ be the oracle Turing machine from Theorem 2.3. For all $S \in \mathbf{C}$ and $\epsilon > 0$, there is an oracle Turing machine $M_e$ and a sequence $R' \in \mathbf{C}$ such that*

1. *$S \leq_{\mathrm{wtt}} R'$ via $M_d$ and $R' \leq_\mathrm{T} S$ via $M_e$.*

2. *$\rho^-_{M_d}(S, R') \leq \dim_\mathrm{H}(S) + \epsilon$.*

3. *$\rho^+_{M_d}(S, R') \leq \dim_\mathrm{P}(S) + \epsilon$.*

**Proof.** Let $S \in \mathbf{C}$ and choose some sequence $R$ for $S$ as in Theorem 2.3. Let $\delta = \epsilon/4$. Let $D \in (\dim_\mathrm{P}(S) + 3\delta, \dim_\mathrm{P}(S) + 4\delta)$ and $d \in (\dim_\mathrm{H}(S), \dim_\mathrm{H}(S) + \delta)$ both be rational. By Theorem 2.3, there exists $n_0 \in \mathbb{N}$ such that, for all $n \geq n_0$, $\#(M_d^R, S[0\mathinner{.\,.}n-1]) < Dn$.

$M_e$ will make use of the oracle Turing machine $M_d$. The proof of Theorem 2.3 in [4] shows that $M_d$ has the following useful properties. First, write $S = s_1 s_2 s_3 \ldots$ and $R = r_1 r_2 r_3 \ldots$, where each $s_i, r_i \in \{0,1\}^*$ are blocks such that $|s_i| = i$ and $|r_i| \leq |s_i| + o(|s_i|)$.

- $M_d$ computes $S$ from $R$ in stages, where it outputs the block $s_i$ on the $i^\text{th}$ stage.

- Assuming that $M_d$ has already computed $s_1 \ldots s_i$, $M_d$ uses only the block $r_{i+1}$ and the prefix $s_1 \ldots s_i$ to compute $s_{i+1}$.

Because of these properties, we can use $M_d$ to search for a sequence $R'$ that satisfies requirements 1, 2 and 3 in the statement of Theorem 2.4, as well as the auxiliary requirement stated at the start of this proof. By Theorem 2.3, $R$ satisfies these requirements, so such an $R'$ will exist. By the above two properties of $M_d$, if we find a string $r' = r'_1 \ldots r'_i$ that satisfies requirements 1, 2 and 3 (in the sense described below, where $r'_i$ is the $i^\text{th}$ block of $R'$, the block $M_d$ reads from $R'$ when outputting $s_i$), we will always be able to find an extension $r'' = r'_{i+1} \ldots r'_j$ (for some $j > i$) such that $r'r''$ continues to satisfy the requirements. It will not matter if $r' \not\sqsubset R$, since $M_d$ does not use the portion of $R$ coming before block $r_{i+1}$ to compute $s_{i+1}$. In other words, to reverse the computation of $M_d^{R'}$ and compute $R'$ from $S$, we don't need to find the $R$ from Theorem 2.3; we need only to find an $R'$ that is "close enough" in terms of query usage. Assume without loss of generality that for any block $r'_k$ we consider, the rightmost bit of $r'_1 \ldots r'_k$ queried by $M$ when computing $s_1 \ldots s_k$ is the last bit of $r'_k$. Then we have that $\#(M_d^{r'_1 \ldots r'_k}, s_1 \ldots s_k) = |r'_1 \ldots r'_k|$.

Define the oracle Turing machine $M_e$ with oracle $S \in \mathbf{C}$ as follows. Let $i \in \mathbb{N}$ and assume inductively that the prefix $r' = r'_1 \ldots r'_i \sqsubset R'$ has been computed, so that, letting $|s_1 \ldots s_i| = n$,

(a) $M_d^{r'}(n)$ outputs $S[0\mathinner{.\,.}n-1]$,

(b) $\#(M_d^{r'}, S[0\mathinner{.\,.}n-1]) \leq dn$,

(c) for all $m$ with $n_0 \leq m \leq n$, $\#(M_d^{r'}, S[0\mathinner{.\,.}m-1]) \leq Dm$.

$M_e^S$ searches all strings $r'' \in \{0,1\}^+$ until it finds one that satisfies, letting $N > n$ be the length of the largest output $M_d^{r'r''}$ can produce without querying beyond its finite oracle string,

(a) $M_d^{r'r''}(N)$ outputs $S[0..N-1]$,

(b) $\#(M_d^{r'r''}, S[0..N-1]) \leq dN$,

(c) for all $m$ with $n_0 \leq m \leq N$, $\#(M_d^{r'r''}, S[0..m-1]) \leq Dm$.

$M_e^S$ then outputs $r''$ and saves it for the computation of the next extension of $R'$. It remains to show that such an $r''$ can always be found.

By the existence of $R$ from Theorem 2.3 and a simple induction on the stages of computation that $M_e$ performs, $M_e^S$ will always be able to find an $r''$ satisfying conditions (a) and (b). It remains to show that among all such $r''$, at least one must satisfy (c).

By Theorem 2.3 and our choice of $D$, for all sufficiently large $k \in \mathbb{N}$, $|r_1 \ldots r_k| \leq (D - 3\delta)|s_1 \ldots s_k|$. How large $k$ must be depends on $D$ and $\delta$ but not on $i$ (the number of blocks satisfying the inductive hypothesis), so assume that $i$ is larger than this value. Then for all $k > i$,

$$
\begin{aligned}
&|r'_1 \ldots r'_i r_{i+1} \ldots r_k| \\
&= |r_1 \ldots r_k| + |r'_1 \ldots r'_i| - |r_1 \ldots r_i| \\
&\leq (D - 3\delta)|s_1 \ldots s_k| + |r'_1 \ldots r'_i| - |r_1 \ldots r_i| && \text{Theorem 2.3 and our choice of } D \\
&\leq (D - 3\delta)|s_1 \ldots s_k| + d|s_1 \ldots s_i| - |r_1 \ldots r_i| && \text{inductive hypothesis (b)} \\
&\leq (D - 3\delta)|s_1 \ldots s_k| + d|s_1 \ldots s_i| - (\dim_H(S) - \delta)|s_1 \ldots s_i| && \text{[4, Lemma 4.1]} \\
&\leq (D - 3\delta)|s_1 \ldots s_k| + (\dim_H(S) + \delta)|s_1 \ldots s_i| - (\dim_H(S) - \delta)|s_1 \ldots s_i| && \text{choice of } d \\
&= (D - 3\delta)|s_1 \ldots s_k| + 2\delta|s_1 \ldots s_i| \\
&\leq (D - 3\delta)|s_1 \ldots s_k| + 2\delta|s_1 \ldots s_k| \\
&= (D - \delta)|s_1 \ldots s_k|. && (1)
\end{aligned}
$$

Let $r'' = r_{i+1} \ldots r_k$. Because each block length is asymptotically smaller than the length of the smallest prefix containing the block ($|s_i| = i$ implies that $|s_1 \ldots s_i| \approx i^2$), it suffices that we have just verified (c) (with "an extra $\delta$ of room") only for $m$ corresponding to boundaries between blocks. To formally justify this, recall that for all $i \in \mathbb{N}$, $|s_i| = i$. Let $m \in \mathbb{N}$, and let $k_m \in \mathbb{N}$ be the largest value of $k \in \mathbb{N}$ for which $s_1 \ldots s_k \sqsubseteq S[0..m-1]$; i.e., the block of $S$ immediately before the block containing $S[m-1]$. Then for all $n_0 \leq m \leq N$,

$$
\begin{aligned}
\#(M_d^{r'_1 \ldots r'_i r_{i+1} \ldots r_{k_m+1}}, S[0..m-1]) &\leq |r'_1 \ldots r'_i r_{i+1} \ldots r_{k_m}| + |r_{k_m+1}| \\
&\leq |r'_1 \ldots r'_i r_{i+1} \ldots r_{k_m}| + |s_{k_m+1}| \\
&= |r'_1 \ldots r'_i r_{i+1} \ldots r_{k_m}| + k_{m+1} \\
&\leq |r'_1 \ldots r'_i r_{i+1} \ldots r_{k_m}| + 2\sqrt{m+1} \\
&\leq (D - \delta)|s_1 \ldots s_{k_m}| + 2\sqrt{m+1} && \text{by inequality (1)} \\
&\leq (D - \delta)m + 2\sqrt{m+1} \\
&\leq Dm.
\end{aligned}
$$

It follows that there exists at least one $r''$ ($= r_{i+1} \ldots r_k$, although $M_e$ may find another $r''$ first) satisfying (a), (b), and (c). Therefore the sequence $R'$ will satisfy requirements 1, 2, and 3 of Theorem 2.4. □

The following theorem is the main result of this paper. It states that constructive packing dimension can be almost optimally extracted from a sequence of positive packing dimension, while at the same time, constructive Hausdorff dimension is *partially* extracted from this sequence, if it has positive Hausdorff dimension and packing dimension less than 1. The machine $M_e$ from Theorem 2.4 serves as the extractor. Intuitively, this works because $M_e$ compresses the sequence $S$ into the sequence $R$. Since $R$ is a compressed representation of $S$, $R$ must itself be more incompressible than $S$. However, because dimension measures the compressibility of a sequence, this means that the constructive dimensions $R$ are greater than those of $S$.

**Theorem 2.5.** *For all $\epsilon > 0$ and $S \in \mathbf{C}$ such that $\dim_P(S) > 0$, there exists $R \equiv_T S$ such that $\dim_P(R) \geq 1 - \epsilon$ and $\dim_H(R) \geq \frac{\dim_H(S)}{\dim_P(S)} - \epsilon$.*

**Proof.** Let $\epsilon > 0$ and $S \in \mathbf{C}$ such that $\dim_P(S) > 0$. Let $\delta > 0$ and $R', M_d$ be as in Theorem 2.4. Let $R'' \in \mathbf{C}$ and $M \in \text{OTM}$ such that $R' \leq_T R''$ via $M$, $\rho_M^-(R', R'') = \dim_H(R')$ and $\rho_M^+(R', R'') = \dim_P(R')$ (the existence of $M$ and $R''$ is asserted by Theorem 2.3). By Lemma 1.1, we have
$$\rho^+(S) \leq \rho_{M_d}^+(S, R')\rho_M^+(R', R''),$$
which, by construction of $R'$ and $R''$ implies $\rho^+(S) \leq (\dim_P(S)+\delta)\dim_P(R')$. Since $\rho^+(S) = \dim_P(S)$,
$$\dim_P(R') \geq \frac{\dim_P(S)}{\dim_P(S) + \delta}.$$

Moreover (by Lemma 1.1 again), $\rho^-(S) \leq \rho_{M_d}^+(S, R')\rho_M^-(R', R'')$, which, by construction of $R'$ and $R''$, implies $\rho^-(S) \leq (\dim_P(S) + \delta) \dim_H(R')$. Since $\rho^-(S) = \dim_H(S)$,
$$\dim_H(R') \geq \frac{\dim_H(S)}{\dim_P(S) + \delta}.$$

Taking $\delta$ small enough, we get by the above inequalities: $\dim_P(R) \geq 1 - \epsilon$ and $\dim_H(R) \geq \frac{\dim_H(S)}{\dim_P(S)} - \epsilon$. □

Theorem 2.5 has a number of applications, stated in the following corollaries, which shed light on the constructive dimensions of sequences, spans and degrees.

**Corollary 2.6.** *Let $S \in \mathbf{C}$ and assume that $\dim_H(S) > 0$. Then the Hausdorff dimensions $\dim_H(\deg_T(S))$ and $\dim_H(\text{span}_T(S))$ are both at least $\frac{\dim_H(S)}{\dim_P(S)}$.*

We obtain a zero-one law for the constructive packing dimension of Turing lower spans and degrees.

**Corollary 2.7.** *For all $S \in \mathbf{C}$, the packing dimensions $\dim_\mathrm{P}(\deg_\mathrm{T}(S))$ and $\dim_\mathrm{P}(\mathrm{span}_\mathrm{T}(S))$ are each either 0 or 1.*

Because of the extension of Theorem 2.1 to Turing reductions by Joe Miller [15], we must settle for more conditional results for constructive Hausdorff dimension. We focus attention on regular sequences.

**Corollary 2.8.** *For all $\epsilon > 0$ and all regular $S \in \mathbf{C}$ such that $\dim_\mathrm{H}(S) > 0$, there exists $R \equiv_\mathrm{T} S$ such that $\dim_\mathrm{H}(R) \geq 1 - \epsilon$.*

**Corollary 2.9.** *For all regular $S \in \mathbf{C}$ such that $\dim_\mathrm{H}(S) > 0$,*

$$\begin{aligned} \dim_\mathrm{H}(\mathrm{span}_\mathrm{T}(S)) &= \dim_\mathrm{H}(\deg_\mathrm{T}(S)) = \\ \dim_\mathrm{P}(\mathrm{span}_\mathrm{T}(S)) &= \dim_\mathrm{P}(\deg_\mathrm{T}(S)) = 1. \end{aligned}$$

We note that the zero-one law for the constructive packing dimension of Turing and Turing lower spans and degrees also follows from the following theorem due to Fortnow, Hitchcock, Pavan, Vinodchandran and Wang [6], giving a polynomial-time extractor for constructive packing dimension. For $R, S \in \mathbf{C}$, write $R \leq_\mathrm{T}^\mathrm{P} S$ if $R \leq_\mathrm{T} S$ via an OTM that, on input $n$, runs in time polynomial in $n$, and similarly for $\equiv_\mathrm{T}^\mathrm{P}$.

**Theorem 2.10** (Fortnow, Hitchcock, Aduri, Vinodchandran and Wang [6]). *For all $\epsilon > 0$ and all $S \in \mathbf{C}$ such that $\dim_\mathrm{P}(S) > 0$, there exists $R \equiv_\mathrm{T}^\mathrm{P} S$ such that $\dim_\mathrm{P}(R) \geq 1 - \epsilon$.*

In fact, Theorem 2.10 holds for any resource-bounded packing dimension [11] defined by Turing machines allowed at least polynomial space, which includes constructive packing dimension as a special case.

## 3 Nonexistence of Universal Extractors

The Turing reduction in the proof of Theorem 2.5 is uniform in the sense that, for all $\epsilon > 0$, there is a *single* Turing reduction $M$, universal for $\epsilon$ and all sequences $S$, such that $\dim_\mathrm{H}(M(S)) \geq \dim_\mathrm{H}(S)/\dim_\mathrm{P}(S) - \epsilon$.

We can show that there is no *universal* Turing reduction that is guaranteed to increase – to a fixed amount – the dimension of all sequences of sufficiently large dimension.

**Theorem 3.1.** *For every Turing reduction $M$ and all reals $\alpha, \beta$ with $0 < \alpha < \beta < 1$, there exists $S \in \mathbf{C}$ with $\dim_\mathrm{H}(S) \geq \alpha$ such that $M(S)$ does not exist or $\dim_\mathrm{H}(M(S)) < \beta$.*

**Proof.** For this proof, it will be convenient to say that $R \leq_\mathrm{T} S$ via $M$ if $M^S(n)$ outputs $R[n]$, rather than $R[0 \mathinner{.\,.} n-1]$, bearing in mind that both definitions of a Turing reduction are equivalent.

Suppose for the sake of contradiction that there exist real numbers $\alpha, \beta$ with $0 < \alpha < \beta < 1$ and a Turing reduction $M$ such that, for all $S \in \mathbf{C}$ satisfying $\dim_\mathrm{H}(S) \geq \alpha$, then $\dim_\mathrm{H}(R) \geq \beta$, where $R = M(S)$. Fix rationals $\alpha', \gamma$ such that $\alpha < \alpha' < \gamma < \beta$. We will convert $M$ into a tt reduction $N$ (meaning that $N$ is also a wtt reduction) that guarantees the slightly weaker condition that if $\dim_H(S) > \alpha'$, then $\dim_H(N(S)) \geq \beta$. Then for any $S \in \mathbf{C}$ such that $\dim_H(S) = \gamma > \alpha'$, it follows that $\dim_H(N(S)) \geq \beta > \gamma = \dim_H(S)$, which contradicts Theorem 2.1.

On input $n \in \mathbb{N}$ and with oracle sequence $S$, $N^S(n)$ simulates $M^S(n)$. In parallel, for all integers $m > n$, $N$ searches for a program of length at most $\alpha'm$ computing $S[0\mathinner{\ldotp\ldotp}m-1]$. If $N$ finds such a program before the simulation of $M^S(n)$ terminates, then $N$ outputs 0. If instead the simulation of $M^S(n)$ halts before such a short program is found, then $N$ outputs $R[n]$, the output bit of $M^S(n)$.

If $\dim_H(S) < \alpha'$, then for infinitely many $m \in \mathbb{N}$, $C(S[0\mathinner{\ldotp\ldotp}m-1]) \leq \alpha'm$. Therefore $N^S$ halts, although the output sequence $N(S)$ may contain a lot of 0's, which is acceptable because we do not care what $N$ outputs if $\dim_H(S) < \alpha'$.

If $\dim_H(S) \geq \alpha'$, then by hypothesis, $M^S$ is guaranteed to halt and to compute $R$ such that $\dim_H(R) \geq \beta$. Therefore $N^S$ halts, establishing that $N$ is a tt reduction. If $\dim_H(S) = \alpha'$, then once again, we do not care what $N$ outputs. If $\dim_H(S) > \alpha'$, then only finitely many $m$ satisfy $C(S[0\mathinner{\ldotp\ldotp}m-1]) \leq \alpha'm$. Therefore the parallel search for short programs will never succeed once $N$ begins checking only prefixes of $S$ of sufficiently large length. This means that from that point on, $N$ will simulate $M$ exactly, computing a sequence $R'$ that is a finite variation of $R$. Since dimension is unchanged under finite variations, $\dim_H(R') = \dim_H(R) \geq \beta$. □

Theorem 3.1 tells us that, contrary to the proofs of Theorems 2.4 and 2.5, any extractor construction for Turing reductions must make use of some property of the sequence beyond a simple bound on its dimension.

## 4 Bounded Reductions

In this section, we show that bounded reductions cannot extract dimension, and we exhibit sequences from which bounded reductions are able to extract dimension, but not wtt reductions, and vice versa. We also show that any sequence that does not tt-compute a sequence of positive packing dimension cannot bT-compute a sequence of positive Hausdorff dimension.

### 4.1 Nonexistence of Bounded Extractors

The next theorem shows that bounded Turing reductions cannot extract dimension.

**Theorem 4.1.** *For every rational number $\alpha$ with $0 < \alpha < 1$, there exists a sequence $S \in \mathbf{C}$ such that, for all bounded reductions $M$, $\dim_H(M(S)) \leq \dim_H(S) = \alpha$.*

**Proof.** For brevity, we refer to the proof of Theorem 2.1 in [16], as our proof is obtained by a slight modification of that proof. Nies and Reimann in [16] construct $S$ via a finite-injury argument, in which $S$ is a limit-recursive (that is, $\Delta_2^0$) sequence given by the prefixes

$$s_0, s_1, \ldots \in \{0,1\}^*,$$

where, for each $i \in \mathbb{N}$, $m_i = |s_i|$, $s_i \sqsubset s_{i+1}$, and $s_i$ is chosen to diagonalize against the $i^{\text{th}}$ wtt reduction. The finite-injury construction may change each $s_i$ finitely often, thereby injuring all $s_j$ for $j \geq i$, but each change will be to a string $s_i'$ of the same length $m_i$.

The intuitive idea behind the proof of Theorem 2.1 is that each $s_i \in \{0,1\}^{m_i}$ is chosen to be an $\alpha$-incompressible string (that is, $\mathrm{C}(s) \geq \alpha|s|$ for all prefixes $s \sqsubseteq s_i$) that guarantees that the $i^{\text{th}}$ wtt reduction with oracle $s_i$ either diverges or outputs an $\alpha$-compressible string. In parallel, the construction searches for a program of length at most $\alpha|s|$ that outputs a prefix $s \sqsubseteq s_i$, which would contradict the supposed incompressibility of $s_i$. If no such short program is ever found, then this means that $s_i$ is "doing its part" to ensure that $\dim_{\mathrm{H}}(S) \geq \alpha$. If such a short program is found, then a new $s_i' \in \{0,1\}^{m_i}$ is chosen (in order to guarantee that $\dim_{\mathrm{H}}(S) \geq \alpha$), and this new choice of $s_i'$ constitutes an injury to all later prefixes $s_{i+1}, s_{i+2}, \ldots$, a new search must begin for strings extending $s_i'$ rather than $s_i$. The above description is slightly inaccurate; to be more precise, $s_i$ is chosen not only to be $\alpha$-incompressible *itself*, but also to have a measure of at least $2^{-2i-1}$ of $\alpha$-incompressible *extensions*, to ensure future requirements have many extensions of $s_i$ from which to choose. The abundance of incompressible strings ensures that this invariant may be maintained for all $i \in \mathbb{N}$. However, the idea is the same: enumerating too many $\alpha$-compressible extensions of $s_i$ means the $i^{\text{th}}$ requirement must change to a different $s_i'$, causing injury to future extensions.

The key idea needed by Nies and Reimann is that a very large fraction of $\{0,1\}^{m_i}$ is incompressible, which is in turn used to obtain compression of the *output* $M(S)$ (if it exists for wtt reduction $M$). To show that bounded reductions cannot extract dimension, we modify the proof of Theorem 2.1 slightly to introduce one additional type of injury: whenever it is discovered that, for some $n \in \mathbb{N}$, the $n^{\text{th}}$ binary string (call this $\sigma_n$) is computed by a program of length less than $\log \log n$, then we set $S[n] = 0$ (thereby possibly causing an injury in all $s_i$'s of length at least $n$, all of which are now required to obey $s_i[n] = 0$ to ensure that $S[n] = 0$ in the limit). The scarcity of programs of length $< \log \log n$ ensures that only a logarithmic density of such $n$'s exist in any prefix of $S$, so $\dim_{\mathrm{H}}(S)$ remains unchanged. Furthermore, this scarcity ensures that the set of strings of length $m$ that have the bit 0 at all indices $n < m$ such that $\mathrm{C}(\sigma_n) < \log \log n$ is of cardinality at least $2^{m-\log m}$. Since incompressible strings are similarly abundant, we modify the proof of Nies and Reimann so that each $s_i$ is required *both* to have a large measure of extensions that are incompressible *and* that are 0 at every index $n$ with $\mathrm{C}(\sigma_n) < \log \log n$, knowing that there are a large measure of strings satisfying both conditions. In particular, define

$$\mathcal{P} = \{\, Z \in \mathbf{C} \mid \forall n > n_0 \; [\mathrm{C}(Z[0..n-1]) \geq \lfloor \alpha n \rfloor \text{ and } Z[m] = 0 \text{ if } \mathrm{C}(\sigma_m) < \log |\sigma_m|]\,\}.$$

where $n_0$ is large enough that $\mathcal{P}$ has measure at least $1/2$. $S$ is chosen to be an element of $\mathcal{P}$, as in [16]. There, Nies and Reimann [16] defined $\mathcal{P}$ to ensure only the first condition, but it is not difficult to verify that the remainder of their proof goes through with $\mathcal{P}$ as defined above; that is, that the $S$ chosen from $\mathcal{P}$ wtt-computes no sequence of higher Hausdorff dimension than $S$. The key idea is that, in the proof of Lemma 3.2 of [16], if a string $s_i$ is injured by our additional requirement (that is, because $s_i[n] = 1$ initially, but it is later discovered that $\mathrm{C}(n) < \log \log n$), then this discovery must result from enumerating program of length at most $\log \log n < \alpha |s_i|$, which is all that is needed for the original proof to work.

It remains to show that this modification ensures that no bounded reduction can extract Hausdorff dimension from $S$. Let $M$ be a bounded reduction and let $R = M(S)$, if it exists, and let $k \in \mathbb{N}$. Then, whenever $M^S(k)$ queries $S[n]$ for some $n \in \mathbb{N}$, this constitutes a computation of $\sigma_n$ from an input of length $\log k + O(1)$, implying that $\mathrm{C}(\sigma_n) \leq \log k + O(1)$. (The extra $O(1)$ information is needed to specify $M$, as well as the order in which the query is made and the answers to queries prior to $\sigma_n$, both of which are constant length if $M$ is a bounded reduction.)

If $n > 2^{2^{2^k}}$, then the answer to the query must be 0, since we ensured that all $n$ computable from an input as short as $k$ satisfy $S[n] = 0$. Therefore, on input $k$, any query beyond $2^{2^{2^k}}$ may be replaced by a constant 0 to obtain the *wtt* reduction $M'$, which will behave the same as $M$, implying $R = M'(S)$. But Theorem 2.1 tells us that $M'$ cannot extract dimension from $S$ because it is a wtt reduction, so $\dim_{\mathrm{H}}(R) \leq \alpha$. $\square$

## 4.2 Bounded Reductions versus Weak Truth-Table Reductions

The next theorem exhibits a sequence from which a bounded reduction can extract packing dimension, but no weak truth-table reduction can extract packing dimension.

**Theorem 4.2.** *There is a sequence $A \in \mathbf{C}$ such that every sequence $B \leq_{\mathrm{wtt}} A$ satisfies $\dim_{\mathrm{P}}(B) = 0$ while there exists a sequence $E \leq_{\mathrm{bT}} A$ that satisfies $\dim_{\mathrm{P}}(E) = 1$.*

**Proof.** For a recursively enumerable language $R$ with some standard enumeration $R_0, R_1, R_2, \ldots$, we define the convergence-module $c_R$ of $R$ by

$$c_R(x) = x + \min \{ \ s \in \mathbb{N} \ | \ \forall y \leq x \ [R_s[y] = R[y]] \ \}$$

Notice that $R$ computes $c_R$ and that any function that majorizes $c_R$ computes $R$. Let now $L$ be a recursively enumerable maximal language of complete Turing-degree (see [18] for definitions). The maximality of $L$ ensures that the complement $\bar{L}$ of $L$ is very sparse in the sense that the function $n \mapsto$ (the $n$-th integer that does not belong to $L$) dominates every recursive function. Let $\Omega$ be Chaitin's Martin-Löf random sequence [3] and let $B$ be a recursively enumerable language of high but incomplete Turing degree with the property that $c_B$ dominates every recursive function. Such sequences $B$ exist [25, Page 220, Exercise XI.2.7]. Now define the following language $A$:

$$\langle x, y \rangle \in A \iff (x \in L \wedge y = 0) \vee (x \in \Omega \wedge y > c_B(x)).$$

For the first part of the theorem, we show that every sequence wtt-reducible to $A$ has packing dimension 0. For this, it is enough to show that for every computable function $f$ and almost all $n$ the plain Kolmogorov-complexity of $A[0 \mathinner{\ldotp\ldotp} f(n)-1]$ is of order logarithmic in $n$ and hence no wtt-reduction with use $f$ can produce a sequence of positive packing dimension. Now, given $f$ computable (which we can assume to be increasing), note that $c_B$ dominates the function $m \mapsto f(2^m)$. Furthermore, there are, for sufficiently large $n$, at most $\log n$ non-elements of $L$ below $f(n)$. Moreover, for sufficiently large $n$, if a pair $\langle x, y \rangle \leq f(n)$ (which in particular means that both $x$ and $y$ are smaller than $f(n)$) satisfies $y \geq c_B(x)$, then one has $f(n) \geq y \geq c_B(x) \geq f(2^x)$, which, since $f$ is increasing implies $x \leq \log n$. Thus, for sufficiently large $n$, a pair $\langle x, y \rangle \leq f(n)$ is in $A$ if and only if either $x \in L \wedge y = 0$ or $x \leq \log n \wedge y > c_B(x) \wedge x \in \Omega$. Hence, for almost $n$, $C(A[0 \mathinner{\ldotp\ldotp} f(n)-1]) \leq 4 \log n$: one can use $\log n$ bits to code $B$ below $\log n$, one can use $\log n$ bits to code $\Omega[0 \mathinner{\ldotp\ldotp} \log n - 1]$ and one can use $\log n$ bits to code how many elements of $L$ below $f(n)$ are not in $L$; then enumerating $L$ long enough gives away which of the elements below $f(n)$ are in $L$ and which not. The remaining $\log n$ bits are an overhead used to absorb the various constants stemming from coding this information within the framework of a given universal machine and so on. These arguments show then that no sequence of positive packing dimension is wtt-reducible to $A$.

For the second part of the theorem, we construct a sequence $E \leq_{\text{bT}} A$ which has packing dimension 1. For this, let $I_n = \{n!, n!+1, n!+2, \ldots, (n+1)!-1\}$ and define $E[x]$ for $x \in I_n$ as follows: First check whether $\langle n, 0 \rangle \in A$. If not, let $E[x] = 0$. If so, determine the time $y$ needed to enumerate $n$ into $L$ and let $E[x] = A[\langle x, y \rangle]$. This algorithm shows that $E[x]$ can be computed relative to $A$ with 2 queries and $E \leq_{\text{bT}} A$. As $L \not\leq_T B$ (because $B$ has incomplete r.e. degree) there are infinitely many $n$ such that the time $y$ to enumerate $n$ into $L$ satisfies $y > c_B((n+1)!)$ (if this was not the case, then the function $n \mapsto (n+1)! + c_B((n+1)!)$, which is $B$-recursive, would be an upper bound for $c_L$, which would imply $L \leq_T B$, a contradiction). It follows that for these $n$ the characteristic functions of $\Omega$ and $E$ coincide on $I_n$; hence $C(E[0 \mathinner{\ldotp\ldotp} (n+1)!-1]) \geq \frac{n}{n+2} \cdot (n+1)!$ for almost all these $n$. So $\dim_P(E) = 1$. $\square$

The next theorem exhibits a sequence from which a weak truth-table reduction can extract Hausdorff dimension, but no bounded reduction can even extract packing dimension.

**Theorem 4.3.** *There is a sequence $A \in \mathbf{C}$ such that every sequence $E \leq_{\text{bT}} A$ satisfies $\dim_P(E) = 0$ while there exists a sequence $B \leq_{\text{wtt}} A$ that satisfies $\dim_H(B) = 1$.*

**Proof.** Let $B = \Omega$ be Chaitin's Martin-Löf random sequence and define $A \subseteq \mathbb{N}$ such that

$$\langle x, y \rangle \in A \iff x \in \Omega \wedge C(y) \geq 2^x.$$

It holds that $\Omega \leq_{\text{wtt}} A$ as $x \in \Omega$ if and only if there is a $y \leq 2^x$ with $\langle x, y \rangle \in A$. As $\Omega$ is Martin-Löf random, $\dim_H(\Omega) = 1$, and the second part of the theorem is satisfied.

Now consider any sequence $E \leq_{\text{bT}} A$. For every $n \in \mathbb{N}$, each of the elements queried by the given bT-reduction to compute $E[0 \mathinner{\ldotp\ldotp} n-1]$ have Kolmogorov complexity at most $c \cdot \log n$, for some constant $c \in \mathbb{N}$, by the same argument in the proof of Theorem 4.1. Let $\langle x, y \rangle$ be a query in this computation. As $C(y) \geq 2^x$ if $\langle x, y \rangle \in A$, this query can get an answer "yes" only if $2^x \leq c \cdot \log n$. There are only $c \cdot \log \log n$ many $x$ with this property, and for each such $x$, one can code, using $O(\log n)$ bits, the number of $y$ such that $C(y) < 2^x$. By simulating the universal machine long enough, one may compute all these $y$ for each such $x$. Since there are at most $c \cdot \log \log n$ such $x$ with this property, the set of all queries $\langle x, y \rangle$ such that, when computing $E[0 \mathinner{.\,.} n-1]$, the query $\langle x, y \rangle$ has answer "yes", can be computed from an input of $O(\log n \cdot \log \log n)$ bits.

Hence one can compute $E[0 \mathinner{.\,.} n-1]$ from an input of size $O(\log n \cdot \log \log n)$ by computing the pairs $\langle x, y \rangle$ described above, and then running the bT-reduction with this information to determine which queries have answer "yes". Since this holds for every $n \in \mathbb{N}$, it follows that $\dim_P(E) = 0$. □

## 4.3 Truth-Table and Bounded Reductions

Our final result shows that any sequence that does not tt-compute a sequence of positive *packing* dimension cannot bT-compute a sequence of positive *Hausdorff* dimension.

**Theorem 4.4.** *Let $A \in \mathbf{C}$ such that, for every sequence $E \leq_{tt} A$, $\dim_P(E) = 0$. Then, for every sequence $B \leq_{bT} A$, $\dim_H(B) = 0$.*

**Proof.** Let $k$ be the maximum number of queries made by the bounded Turing reduction $M$ from $B$ to $A$. Furthermore, let $I_n$ contain all numbers $m$ with $n! \leq m < (n+1)!$. Now let $c_n$ be the number of pairs $(m, b_1 b_2 \ldots b_k)$ such that $m \in I_n$, $b_1, b_2, \ldots, b_k \in \{0, 1\}$ and $M$ on input $m$ converges provided that the oracle queries are answers with $b_1, b_2, \ldots, b_k$ where not all bits might be needed. The sequence $c_n / |I_n|$ oscillates between 0 and $2^k$.

Now let $\epsilon$ be any positive real number. One can choose a number $\delta$ with $0 < \delta < \epsilon/4$ such that for almost all $m$ the Kolmogorov complexity of every string in $\{0, 1\}^m$ containing at most $\delta \cdot m$ many digits 1 is at most $m \cdot \epsilon/4$. There is a rational number $q$ such that $q \leq c_n/|I_n|$ for almost all $n$ and $c_n/|I_n| < q + \delta$ for infinitely many $n$. There is a recursive function $f$ such that $f(n)$ bounds for every $n$ with $c_n/|I_n| \geq q$ the convergence time and places of queries for computations of $M$ for $q \cdot |I_n|$ many tuples $(m, b_1, b_2, \ldots, b_k)$ with $m \in I_n$ and $b_1, b_2, \ldots, b_k \in \{0, 1\}$ such that $M$ takes input $m$ and receives the oracle answers taken from $(b_1, b_2, \ldots, b_k)$.

Having this, one can define the following sequence $E$: for all $n$ and all $m \in I_n$, one follows the activity of $M$ on oracle $A$ for $f(n)$ steps and let $E[m]$ be the result provided that $M$ terminates within the given time $f(n)$ and that $M$ also does not query $A$ beyond $f(n)$. Otherwise $E[m] = 0$.

This is a truth-table reduction and hence the packing dimension of $E$ is 0. Hence $C(E[0 \mathinner{.\,.} (n+1)!]) \leq |I_n| \cdot \delta$ for almost all $n$. By construction $E \subseteq B$ and $|I_n \cap (B - E)| \leq \delta \cdot |I_n|$ for the $n$ with $q \leq c_n/|I_n| < q + \delta$. Furthermore, $n!/|I_n| < \delta \cdot |I_n|$ for almost all $n$ and so it follows that the conditional Kolmogorov complexity of $B[0 \mathinner{.\,.} (n+1)!]$ given $E[0 \mathinner{.\,.} (n+1)!]$ is at most $2 \cdot \delta \cdot (n+1)!$ for infinitely many $n$. It follows therefore that $C(B[0 \mathinner{.\,.} (n+1)!]) \leq \epsilon \cdot (n+1)!$ for infinitely many $n$. As the choice of $\epsilon > 0$ was arbitrary, $\dim_{\mathrm{H}}(B) = 0$. $\square$

**Acknowledgments.** We thank Joe Miller for assistance with the proof of Theorem 2.4, as well as John Hitchcock, Jan Reimann and André Nies for their insightful comments. Chris Conidis was very helpful in finding problems with the original proof of Theorem 2.4. We also thank the American Institute of Mathematics which generously invited us to the Workshop on Effective Randomness; this paper is a result of a workgroup discussing open questions during this workshop. Besides the American Institute of Mathematics, we would also like to thank the organizers Denis Hirschfeldt and Joe Miller of the workshop as well as the participants who discussed this research topic with us.